\documentclass[conference]{IEEEtran}
\IEEEoverridecommandlockouts
    
\def\BibTeX{{\rm B\kern-.05em{\sc i\kern-.025em b}\kern-.08em
    T\kern-.1667em\lower.7ex\hbox{E}\kern-.125emX}}

\usepackage[utf8]{inputenc}
\usepackage[algo2e,linesnumbered,ruled,vlined]{algorithm2e}
\usepackage{graphicx}
\usepackage{float}
\usepackage{subcaption}
\usepackage{booktabs}
\usepackage{cancel}
\usepackage{amsmath}
\usepackage{comment}
\usepackage{xcolor}
\usepackage{multirow}
\usepackage{url}
\usepackage{amsfonts}
\usepackage[show]{chato-notes}
\usepackage{enumitem}

\DeclareMathOperator*{\argmin}{arg\,min}

\newcommand{\Nameapproach}[0]{\textsc{PWR}\xspace}

\makeatletter
\newcommand{\linebreakand}{%
  \end{@IEEEauthorhalign}
  \hfill\mbox{}\par
  \mbox{}\hfill\begin{@IEEEauthorhalign}
}
\makeatother

\begin{document}

\title{Power- and Fragmentation-aware\\Online Scheduling for GPU Datacenters}

\author{
\IEEEauthorblockN{
Francesco Lettich, Emanuele Carlini, Franco Maria Nardini, Raffaele Perego, Salvatore Trani
\thanks{This work has been submitted to the IEEE for possible publication. Copyright may be transferred without notice, after which this version may no longer be accessible.}
}
\IEEEauthorblockA{
\textit{Istituto di Scienza e Tecnologie dell'Informazione ``Alessandro Faedo", Consiglio Nazionale delle Ricerche, Pisa, Italy}\\
Email: \{francesco.lettich, emanuele.carlini, francomaria.nardini, raffaele.perego, salvatore.trani\}@isti.cnr.it}
}

\maketitle

\begin{abstract}
The rise of Artificial Intelligence and Large Language Models is driving increased GPU usage in data centers for complex training and inference tasks, impacting operational costs, energy demands, and the environmental footprint of large-scale computing infrastructures.
This work addresses the online scheduling problem in GPU datacenters, which involves scheduling tasks without knowledge of their future arrivals. We focus on two objectives: minimizing GPU fragmentation and reducing power consumption. GPU fragmentation occurs when partial GPU allocations hinder the efficient use of remaining resources, especially as the datacenter nears full capacity. A recent scheduling policy, Fragmentation Gradient Descent (FGD), leverages a fragmentation metric to address this issue.
Reducing power consumption is also crucial due to the significant power demands of GPUs. 
To this end, we propose PWR, a novel scheduling policy to minimize power usage by selecting power-efficient GPU and CPU combinations. This involves a simplified model for measuring power consumption integrated into a Kubernetes score plugin.
Through an extensive experimental evaluation in a simulated cluster, we show how PWR, when combined with FGD, achieves a balanced trade-off between reducing power consumption and minimizing GPU fragmentation.
\end{abstract}


\section{Introduction}
\label{sec:introduction}
The rise of Machine Learning (ML) and large language models (LLMs) significantly amplifies the demand for GPUs in modern large-scale computing infrastructures, commonly called GPU datacenters. Indeed, the adoption of ML and LLMs is expected to rapidly grow further with larger models \cite{wei2022emergent} having higher training and inference costs \cite{snell2024scaling}. 
This, in turn,  drives up operational costs and power consumption of GPU datacenters, thus exacerbating their environmental impact \cite{dodge2022measuring, gao2020smartly, rostirolla2022survey, sg2020new}.
In this work, we address the problem of online scheduling in GPU datacenters handling hybrid ML workloads. This requires scheduling tasks to nodes as they arrive without prior knowledge of future task arrivals.
To optimize GPU utilization, GPU datacenters typically allow multiple tasks to share the same GPU within given resource constraints \cite{weng2022mlaas, weng2023beware, ye2024deep}: this is called \textit{GPU-sharing}, and is realized via policies such as time-sharing or NVIDIA Multi-Process service. 
In this setting, we optimize and reconcile two distinct objectives: \textit{minimizing  GPU fragmentation}, and \textit{minimizing} the \textit{power consumption}.

The first objective, minimizing GPU fragmentation, is directly related to maximize GPU usage. GPU fragmentation occurs when the allocation of partial GPUs prevents the remaining GPU resources from being effectively used. This problem typically arises when a GPU datacenter \textit{operates close} to its overall \textit{capacity}, and manifests in a significant fraction of GPU resources that remain unused. This not only represents a waste of money but also prevents the successful scheduling of additional tasks.
Very recent literature  addresses GPU fragmentation by proposing a novel  fragmentation metric integrated within the Fragmentation Gradient Descent (FGD) scheduling policy \cite{weng2023beware}. This greedy approach leverages prior knowledge of statistics concerning GPU and CPU resources requested by tasks in typical workloads (derived from historical traces) and schedules a task to a node as follows: first, it hypothetically schedules the task to some node, virtually affecting the node's available resources. Secondly, it assesses how much the GPU fragmentation, computed as the expected amount of GPU resources in the node that cannot be allocated to tasks randomly sampled from the target workload, increases compared to not doing the scheduling on that node. Once all the nodes are evaluated, the one with the smallest expected GPU fragmentation increase is selected for scheduling.

The second objective is to reduce the overall power consumption of a GPU datacenter. Their power consumption is massive, primarily due to the high power demands of GPUs, and is expected to increase in the foreseeable future. However, datacenters, on average, do not operate close to their full capacity \cite{hu2021characterization}, thus opening up opportunities to achieve substantial power savings. 
For instance, strategies to minimize power consumption can include consolidating tasks to fully utilize fewer GPUs, thus allowing others to remain in low-power states. Additionally, power-aware scheduling can consider the power profiles of different CPU and GPU models, selecting the most power-efficient combination for a given task.

In this paper, we first introduce a simplified model to measure the power consumption of GPUs and CPUs in a GPU datacenter. We then use this model in a novel scheduling policy named \Nameapproach, made available as a Kubernetes score plugin, that takes scheduling decisions by minimizing the increase in power consumption each time a task is assigned, based on the profiles of CPUs and GPUs available. 
Furthermore, we use \Nameapproach in combination with FGD to strike tradoffs between minimizing GPU fragmentation and minimizing power consumption of a GPU datacenter. The results of reproducible experiments, conducted using real-world traces, highlight that our solution can practically achieve substantial power savings against selected competitors and, for some workloads, up to 20\% versus plain FGD. 


\section{Preliminaries and problem definition}
\label{section:background}

We consider a GPU datacenter managed by the Kubernetes orchestration system and comprising $N$ nodes, which may be heterogeneous in nature. Most of the nodes are equipped with both CPUs and GPUs, though the types of CPUs and GPUs may vary between nodes. However, within any given node, the CPUs and GPUs are of the same type. 
We assume that the datacenter employs some GPU-sharing policy, i.e., multiple tasks are allowed to share a  GPU within the limits of its memory and computational resources. 
%
%
One of these policies is \textit{GPU time-sharing}, 
where a single GPU is shared among multiple tasks by allocating each task a dedicated time slice for exclusive access \cite{weng2023beware}. The main drawback of this approach is the overhead associated with context switching. The other GPU-sharing policy is NVIDIA’s \textit{Multi-Process Service}\footnote{\url{https://docs.nvidia.com/deploy/mps/}}
(MPS) \cite{ye2024deep}, which allows concurrent execution of kernels from different tasks on a single GPU with minimal context-switching overhead by using a shared GPU context. In this work, we assume the use of GPU time-sharing, but the results apply to other GPU-sharing policies such as MPS.
We do not consider NVIDIA’s \textit{Multi-Instance GPU} (MiG) \cite{ye2024deep} a GPU-sharing policy, since it physically partitions the resources of a compatible GPU into isolated instances with their own memory, cache, and compute cores. This slightly reduces overall resource availability, and instances are less power-efficient than the full GPU\footnote{\url{https://kubernetes.web.cern.ch/blog/2023/07/04/efficient-access-to-shared-gpu-resources-part-5/}. Access. 5$^{\text{th}}$ August 2024.}
\cite{li2022characterizing}.

\vspace{0.2em}
\noindent \textbf{Modelling Node Resources.} 
Following the notation provided in \cite{weng2023beware}, we describe the computational resources available on any node $n$ using the node's \textit{unallocated resource vector}: 
$$R_n = \langle R^{CPU}_n, R^{MEM}_n, R^{GPU}_{n, 1}, \dots, R^{GPU}_{n, G_n} \rangle$$ 
In this vector, $R^{CPU}_n$ denotes the number of \textit{virtual unallocated CPUs} (note that $R^{CPU}_n$ is a real number because tasks can require the allocation of fractions of virtual CPUs) and $R^{MEM}_n \ge 0$ is the amount of unallocated RAM. Moreover, given the set of $G_n$ GPUs equipping node $n$, $R^{GPU}_{n, g} \in [0, 1]$ specifies the percentage of unallocated memory and computational resources on a specific GPU $g \in \{1, \ldots, G_n\}$.
In a complementary way, we denote as $Ra_n$ the \textit{allocated resource vector} for node $n$, where: $Ra_n = \langle Ra^{CPU}_n, Ra^{MEM}_n, Ra^{GPU}_{n, 1}, \dots, Ra^{GPU}_{n, G_n} \rangle$.
Symmetrically to $R_n$, each value of $Ra_n$ represents the resources of node $n$ currently allocated to some task.
Finally, we denote by $type_{CPU}(n)$ and $type_{GPU}(n)$ the CPU and GPU model installed in node $n$.

\vspace{0.2em}
\noindent \textbf{Modelling tasks.}
Let $t$ be a task submitted to the datacenter. 
The resources required by the task are described by the  \textit{task resource demand vector} $D_t$ and \textit{task constraint set} $C_t$: 
$$D_t = \langle D^{CPU}_t, D^{MEM}_t, D^{GPU}_t \rangle, \ C_t = \{ C^{CPU}_t, C^{GPU}_t \}$$ 
$D^{CPU}_t$ and $D^{MEM}_t$ are the CPU and memory requests to be satisfied by the corresponding capacities $R_n^{CPU}$ and $R_n^{MEM}$ in $R_n$, while $D^{GPU}_t$ represents the GPU resources required by $t$. 
As in \cite{weng2023beware} we assume that $D^{GPU}_t \in [0,1) \cup \mathbb{Z}^+$, i.e., a task can demand no GPU (0 case), partially use a GPU ($(0,1)$ case), exclusively use one or multiple GPUs ($\mathbb{Z}^+$ case), but cannot demand to both (1) share a GPU with other tasks and (2) fully utilize one or multiple GPUs.
Furthemore, a task might require to be executed on specific CPU or GPU models; if so, such information is indicated in $C^{CPU}_t$ and $C^{GPU}_t$.
Moreover, we assume that if a task is using a fraction of a GPU's resources, and no other tasks are currently using the same GPU, then the task can \textit{opportunistically} use all the computational resources of that GPU while remaining within the memory limit specified by $D_t^{GPU}$.
Finally, we introduce a scalar function $u_n$, used to determine if a node $n$ has sufficient GPU resources to execute some task, i.e. $u_n = \sum_{g = 1}^{G_n} \lfloor R_{n,g}^{\text{GPU}} \rfloor + \max_{g \in \{1,\ldots, G_n\}} (R_{n,g}^{\text{GPU}} - \lfloor R_{n,g}^{\text{GPU}} \rfloor).$

\vspace{0.2em}
\noindent \textbf{Estimating the Power Consumption.} To estimate the power consumption of a GPU datacenter, we limit ourselves to those components that significantly contribute to the overall consumption and have variable power usage,
i.e., the CPUs and GPUs.
Focusing on the CPUs, recall that $R_n^{CPU}$ ($Ra_n^{CPU}$) represents the node's \textit{virtual unallocated  (allocated) CPUs}. 
In modern GPU datacenters,
2 virtual CPUs are typically mapped to a CPU physical core. Thus, $R_n^{CPU}$ can be used to directly determine the number of physical CPUs being currently used in a node.
Let $p_{max}(type_{CPU}(n))$ and $p_{idle}(type_{CPU}(n))$ be functions that return the \textit{maximum} and \textit{minimum} power consumption of a physical core of the specific CPU model present in $n$. Moreover, let also  $ncores(type_{CPU}(n))$ be the number of physical cores present in the CPU type equipping node $n$. 
Then, we estimate the \textit{total power} consumed by the node's CPUs as:
\begin{equation}
\label{eq: cpu pwr cons}
\begin{split}
p&_{CPU}(n) = \\
& p_{max}(type_{CPU}(n)) \cdot \left\lceil \frac{Ra^{CPU}_n}{2 \cdot ncores(type_{CPU}(n))} \right\rceil \ + \\ 
& p_{idle}(type_{CPU}(n)) \cdot \left\lfloor \frac{R^{CPU}_n}{2 \cdot ncores(type_{CPU}(n))} \right\rfloor
\end{split}
\end{equation}
According to this modeling, CPUs that are even minimally used (first term of the sum, with the ceil operator) are assumed to consume maximal power, while those that are idle (second term, with the floor operator) are assumed to consume the lowest possible power. While this formula is bound to overestimate the power consumption of the CPUs of a node, we argue that it provides a safe and reasonable estimation when used in a simulator or when it is not possible to measure the actual power consumption in real-time.
Similarly, for the GPUs of a node, we define $p_{max}(type_{GPU}(n))$ and $p_{idle}(type_{GPU}(n))$ to be the functions that return the \textit{maximum} and \textit{minimum} power consumed by the GPU model present in a node $n$, respectively. Then, we estimate the total power consumed by the $G_n$ GPUs of $n$ as:
\begin{equation}
\label{eq: gpu pwr cons}
p_{GPU}(n) = \sum_{g = 1}^{G_n} 
\begin{cases}
p_{max}(type_{GPU}(n)) \, \text{if } Ra^{GPU}_{n,g} > 0 \\ 
p_{idle}(type_{GPU}(n)) \ \text{otherwise}
\end{cases}
\end{equation}
Again, even if a GPU is not fully occupied, we assume that it is consuming maximum power: this safely covers the assumption that tasks that are using a fraction of a GPU's resources can opportunistically use its free computational resources.

The total power consumption of a node $n$ can be then estimated as $p(n) = p_{CPU}(n) + p_{GPU}(n)$, and the total power consumption of a GPU datacenter  estimated as:
\begin{equation}
\label{eq: datacenter estimated power consumption}
P_{datacenter} = \sum_{n=1}^N p(n)
\end{equation}
\noindent \textbf{GPU Fragmentation.}  Modelling GPU fragmentation requires to first define the concept of \textit{target workload} $M$ derived from historical data \cite{weng2023beware}. A target workload describes how tasks are categorized based on the amount of CPU and GPU resources they require, effectively classifying tasks into different classes.
For example, a target workload $M$ may include a task class $m$ that requests resources $D_m = \langle D_m^{CPU} = 8$, $D_m^{GPU} = 2 \rangle$, and another class that requests $D_{m'} = \langle D_{m'}^{CPU} = 32$, $D_{m'}^{GPU} = 8 \rangle$, among others. Each class $m$ is also associated with a popularity score $p_m$, which represents the \textit{probability} of tasks from that class appearing in the workload $M$.
The GPU fragmentation of node $n$ for class $m$, denoted as $F_n(m)$, measures how many of the node’s unallocated GPU resources cannot be used by a task from class $m$. For further details on the two cases that determine how $F_n(m)$ is computed, we refer the reader to \cite{weng2023beware}.
%
%
%
Once the GPU fragmentation of a node is computed for each task class $m \in M$, the \textit{expected GPU fragmentation of a node $n$}, with respect to a task randomly sampled from the target workload $M$, can be estimated as:
$F_n(M) = \sum_{m \in M} p_m F_n(m)$.
The expected GPU fragmentation $F_n$ is thus a simple statistical measure reflecting how fragmented the GPU resources of node $n$ are, on average, for tasks across all classes in the workload. In turn, the  \textit{expected GPU fragmentation of a datacenter} for a target workload M can be computed as:
\begin{equation}
\label{eq: gpu frag}
F_{datacenter} = \sum_{n=1}^N F_n(M)
\end{equation}
\noindent \textbf{Problem Definition.}
We now consider the online scheduling problem, which involves assigning tasks to nodes as they arrive, without prior knowledge of future task arrivals. 
Each task scheduling is an atomic operation, i.e., a new scheduling decision starts only after the previous one has completed.
The goal is to assign an incoming task $t$ to a suitable node $n \in \{1,\ldots, N\}$. We denote this assignment as $t \rightarrow n$. As before, let $M$ represent the target workload, modeled 
as discussed earlier.
%
The objective is to determine the optimal assignment of task $t$ to the node $n$ of the datacenter that minimizes the following objective function:
\begin{equation}
\label{eq: problem def}
\argmin_{t \rightarrow n} \ \alpha \cdot P_{datacenter} + (1 - \alpha) \cdot F_{datacenter},
\end{equation}
where $\alpha$ is a weighting parameter that balances between the estimated power consumption  and  GPU fragmentation. 
This optimization is subject to the following constraints, assuring that the task's demands do not exceed the resources available in a node:
\begin{equation}
\label{eq: problem def constraints}
\begin{split}
\text{\textbf{(Cond. 1)} } & D_t^{CPU} \leq R_n^{CPU} \\
\text{\textbf{(Cond. 2)} } & D_t^{MEM} \leq R_n^{MEM} \\
\text{\textbf{(Cond. 3)} } & 
\begin{cases}
D_t^{GPU} \leq u_n \text{ if } D_t^{GPU} \in \mathbb{Z}^+ \\
D_t^{GPU} \leq (u_n - \lfloor u_n \rfloor) \, \text{if} \, D_t^{GPU} \in (0,1) 
\end{cases} \\
\text{\textbf{(Cond. 4)} } & \text{if } C_t^{CPU} \neq \emptyset \text{ then } type_{CPU}(n) \in C_t^{CPU} \\
\text{\textbf{(Cond. 5)} } & \text{if } C_t^{GPU} \neq \emptyset \text{ then } type_{GPU}(n) \in C_t^{GPU} \\
\end{split}
\end{equation}

\section{Related work}
\label{section:relatedwork}

In this work, we address the problem of minimizing power consumption in the context of scheduling policies for GPU datacenters dealing with hybrid ML workloads. 
Existing research in this area can be broadly categorized into two approaches: \textit{batch scheduling} and \textit{online scheduling}.
Batch scheduling 
processes batches of tasks at predetermined intervals, while \textit{online scheduling} assigns tasks to nodes as they arrive, without knowledge of future task arrivals.
While batch scheduling often leads to better optimization outcomes, 
online scheduling typically relies on greedy algorithms, which may result in suboptimal solutions. Research focusing on batch scheduling often aims to optimize multiple objectives with integer linear programming, or by using 
evolutionary algorithms, bio-inspired methods, machine learning techniques, or combinations of these \cite{manumachu2017bi, tai2023, liu2020energy, kocot2023energy, ye2024deep}. 
%
Batch scheduling approaches generally rely on several assumptions about task characteristics, such as fixed task arrival times and known resource requirements. 
Consequently, hereinafter we focus on works that address online scheduling, which offers greater adaptability and is more aligned with the objectives of this study.


The recent seminal work by \cite{weng2023beware} introduces a formal framework for characterizing online task scheduling in modern GPU datacenters. Their model assumes that a GPU datacenter must handle hybrid ML workloads and enable GPU time-sharing for maximizing the utilization of GPU resources. Their work does not focus on power consumption, and in this paper we extend their framework to deal effectively with both GPU fragmentation and power consumption. 

Hu et al. \cite{hu2021characterization}  introduce the Cluster Energy Saving (CES) service, which operates within an orchestrator and uses a gradient-boosted decision tree model trained on various node features to predict future cluster behavior. Specifically, the model predicts how many idle nodes can be powered down using Dynamic Resource Sleep (DRS) while maintaining cluster usability. This solution is orthogonal to ours. More broadly, we believe that our contribution, focused on power-aware online scheduling for GPU datacenters, can be combined with hardware-level techniques like DRS or Dynamic Voltage and Frequency Scaling (DVFS) \cite{wu2014green, patel2023towards} to further reduce energy consumption.

In \cite{gao2020smartly}, the authors consider a datacenter powered by variable renewable energy sources (RES) like solar and wind. They partition the datacenter into physical machine (PM) areas and map each to a subset of RESs. Their goal is to online schedule jobs with similar service-level objectives (SLOs) to PM areas that can meet these requirements. To this end, they propose a renewable resource allocation system that predicts the energy production of each RES and the energy demand of each PM area, then matches RESs to PM areas to minimize SLO violations, energy costs, and carbon emissions. They use long short-term memory (LSTM) networks for predictions and a combination of reinforcement learning and linear programming for efficient allocation. Again, the authors' proposal is orthogonal to ours, as it does not exploit energy- or power- aware scheduling techniques, nor it considers the specificities of GPUs and GPU datacenters.

ANDREAS \cite{filippini2021andreas} is an online scheduling solution aimed at reducing the energy consumption of deep learning training workloads. 
When a job is submitted, a Job Profiler collects execution data by running the job on different resource configurations and stores this information in a database. A Job Optimizer uses this profiling data to determine the optimal resource allocation through a randomized greedy heuristic, which balances energy costs and job deadlines. Finally, a Job Manager deploys jobs based on these decisions, handling preemption and migration if needed. Overall, this work does not consider ML hybrid workloads, assumes that training tasks are recurrent and can be effectively profiled, and does not consider the problem of GPU fragmentation or the use of GPU-sharing policies.
Finally, \cite{gu2024greenflow} introduces GreenFlow, a scheduler for GPU datacenters designed to reduce average job completion time while adhering to a given carbon emission budget. GreenFlow focuses exclusively on deep learning training workloads and, more crucially, does not address GPU-sharing, a key aspect of our work. 
%

\textit{In conclusion, to the best of our knowledge our work is the first that addresses power- and GPU fragmentation-aware online task scheduling for GPU datacenters dealing with hybrid ML workloads, with \cite{weng2023beware} being the most related.}

\section{Power-aware Scheduling Policy: \Nameapproach}
\label{section:proposed approach}

As described in Equation \ref{eq: datacenter estimated power consumption} from Section \ref{section:background}, we model power consumption using the estimated power consumption, denoted as $P_{datacenter}$. The pseudocode in Algorithm \ref{alg: PWR} shows how the Kubernetes scheduling framework operates when the score plugin implementing our power-aware scheduling policy \Nameapproach is used.
\begin{algorithm2e}[t]
\footnotesize
\caption{The \Nameapproach scheduling policy}
\label{alg: PWR}
\DontPrintSemicolon
\SetInd{0.5em}{0.5em}
\KwIn{Datacenter nodes $N$, incoming task $t$.}
\KwOut{Assigned node $n^*$}
\SetKwFor{ParallelFor}{parallel for}{do}{end}
\SetKw{Continue}{continue}

\BlankLine
\BlankLine
{
    $S \leftarrow \emptyset$ \nllabel{alg1: init1}\;
    $n^* \leftarrow \bot$ \nllabel{alg1: init2}\;
    \ParallelFor{$n \in N$ \nllabel{alg1: for start}}
    {
        \If{(insufficient resources \textbf{or} unmet constraints)  \nllabel{alg1: filter}}
        {
            \Continue \tcp*[f]{Ignore unsuitable nodes}
        }
        $n^h \leftarrow \textsc{HypAssignToNode}(t, n)$ \nllabel{alg1: hyp assign} \tcp*[r]{Hyp. assignment}
        $\Delta \leftarrow p(n^h) - p(n)$ \nllabel{alg1: delta} \tcp*[r]{Comp. consumption increment}
        $S \leftarrow S \cup (n, \Delta)$ \nllabel{alg1: for end}
    }
    \If{$S \neq \emptyset$ \nllabel{alg1: select node 1}}{
        $n^* \leftarrow \arg \min_{n \in S} \Delta$ \nllabel{alg1: select node 2} \tcp*[f]{Pick node with smallest $\Delta$}
    }
}
\end{algorithm2e}
The scheduler takes in input the task $t$ and the current status of the datacenter, including 
detailed information about its nodes, their specifications, and their (un)allocated resources. 
The variables $S$ and $n^*$ are then initialized (lines \ref{alg1: init1}–\ref{alg1: init2}). Here, $S$ is an initially empty set of pairs, where each pair represents the increase in power consumption if task $t$ were scheduled on a particular node $n$. The variable $n^*$, initially undefined, represents the node on which $t$ will eventually be scheduled.
The \textit{parallel for} loop, between lines \ref{alg1: for start} and \ref{alg1: for end}, evaluates the impact on the datacenter’s power consumption if task $t$ were scheduled on each node in $N$. For each node $n$, the scheduler first checks whether the node has sufficient resources and meets all task constraints (line \ref{alg1: filter}). This check is actually performed by the filter plugin within the Kubernetes scheduling framework.
If node $n$ satisfies both conditions, the scheduler applies the logic of \Nameapproach through the score plugin. It begins by hypothetically assigning task $t$ to node $n$ using the \textsc{HypAssignToNode} function (line \ref{alg1: hyp assign}). This function simulates the assignment by creating an updated copy of $n$’s allocated and unallocated resource vectors, represented by the variable $n^h$. Next, the scheduler calculates the increase in $n$’s total estimated power consumption, denoted as $\Delta$, if task $t$ were to be scheduled on $n$ (line \ref{alg1: delta}). This increase is computed using the power consumption model introduced in Section \ref{section:background}, which takes into account the resource vectors of both $n$ and $n^h$. The result is then stored in $S$ (line \ref{alg1: for end}).
Finally, the Kubernetes scheduling framework assigns task $t$ to the node $n^*$ that meets the resource requirements and results in the smallest increase  $\Delta$ in estimated power consumption, if any (lines \ref{alg1: select node 1}–\ref{alg1: select node 2}).

\subsection{Combining \Nameapproach and \textsc{FGD}: a power- and GPU fragmentation-aware scheduling policy}
Recall that minimizing a GPU datacenter's power consumption is part of the broader problem we defined in Equations \ref{eq: problem def} and \ref{eq: problem def constraints} in Section \ref{section:background}, where minimizing GPU fragmentation is also crucial. Ideally, we aim to combine the benefits of both \Nameapproach and FGD in a single scheduling policy.
Fortunately, the Kubernetes scheduling framework allows to linearly combine the normalized scores of multiple scoring plugins, effectively combining their behaviors provided that an appropriate coefficient $\alpha$ is used. Hence, we use this capability to linearly combine the scores of \Nameapproach with those of FGD.
In the experimental evaluation (Sections \ref{section:experimental settings} and \ref{section:experimental eval}), we explore different $\alpha$ values, identifying those that yield the most effective power- and GPU fragmentation-aware scheduling policies.

\section{Experimental setting}
\label{section:experimental settings}
\label{sec: exp competitors}
We implemented the \Nameapproach scheduling policy as a Kubernetes score plugin written in Go. The plugin is integrated within the customized version of Alibaba's event-driven \textit{open-simulator} used in \cite{weng2023beware}. 
Our code and the related documentation is publicly available in a GitHub repository
\footnote{\url{https://github.com/Fr4nz83/PWR-plugin-kubernetes}}.
We compare our scheduling policy, \Nameapproach, and its combination with FGD, against five heuristic policies that support GPU-sharing and have already been implemented in the simulator:

\begin{enumerate}
\item \textbf{Fragmentation Gradient Descent} (FGD) is the scheduling approach proposed in \cite{weng2023beware}.

\item \textbf{Best-fit} (BestFit) \cite{hadary2020protean} assigns tasks to the node with the least remaining resources, computed as a weighted sum over all resource dimensions.

\item \textbf{Dot-product} (DotProd) \cite{grandl2014multi} allocates tasks to the node with the smallest dot-product between the node’s available resources and the task’s requirements. 

\item \textbf{GPU Packing} (GpuPacking) \cite{weng2022mlaas} prioritizes task assignment first to occupied GPUs, then to idle GPUs on active nodes, and lastly to idle nodes, aiming to preserve resources for multi-GPU tasks.

\item \textbf{GPU Clustering} (GpuClustering) \cite{xiao2018gandiva} packs tasks with similar GPU requirements together to avoid heterogeneous resource distribution on the same node.
\end{enumerate}

\subsection{Traces and generation of workloads}
\label{sec: exp traces}
\begin{table}[h]
    \caption{Distribution of tasks in the Default trace.}
    \centering
    \begin{tabular}{r|c|c|c|c|c|c}
        \textbf{GPU Request per Task} & \textbf{0} & \textbf{(0, 1)} & \textbf{1} & \textbf{2} & \textbf{4} & \textbf{8} \\
        \hline
        \textbf{Task Population (\%)} & 13.3 & 37.8 & 48.0 & 0.2 & 0.2 & 0.5 \\
        \textbf{Total GPU Reqs. (\%)} & 0 & 28.5 & 64.2 & 0.5 & 1.0 & 5.8 \\
        \bottomrule
    \end{tabular}
    \label{tab:task_distribution}
\end{table}
We consider the 2023 Alibaba GPU trace dataset
introduced in \cite{weng2023beware}. 
The dataset includes the \textbf{Default} trace, consisting of 8,152 tasks collected from an Alibaba production-grade GPU datacenter without GPU constraints. The distribution of task profiles and GPU requests in the Default trace is outlined in Table \ref{tab:task_distribution}.
Additionally, from the dataset we consider other three types of traces, which were derived from Default as follows.

\noindent \textbf{Multi-GPU} traces: the amount of GPU resources requested by tasks that use 1 or more entire GPUs is increased by 20\%, 30\%, 40\%, and 50\% compared to the Default trace. This is achieved by increasing the total number of multi-GPU tasks while keeping their internal distribution fixed. The numbers of CPU-only and sharing-GPU tasks remain unchanged.\\
\noindent \textbf{Sharing-GPU} traces: the percentage of GPU resources requested by sharing-GPU tasks is set at 40\%, 60\%, 80\%, and 100\% of the total GPU resources requested by GPU tasks. This is done by adjusting the number of sharing-GPU and multi-GPU tasks, while keeping intra-class distributions fixed and maintaining the same percentage of CPU-only tasks.\\
\noindent \textbf{Constrained-GPU} traces: the percentage of GPU tasks that request specific GPU models is set at 10\%, 20\%, 25\%, and 33\%. All other characteristics match those of Default.

\vspace{0.3em}
\textbf{Workload generation}: to evaluate the datacenter’s capacity under a given scheduling policy, we use a Monte Carlo workload inflation approach for each trace. Tasks are randomly sampled from the traces with replacement and  submitted for scheduling until the cluster reaches full capacity.

\subsection{The simulated GPU datacenter}

The trace data in Section \ref{sec: exp traces} was collected from an Alibaba production-grade GPU datacenter with 1213 nodes, 310 of which do not have any GPU. The datacenter includes a total of 107,018 virtual CPUs and 6,212 GPUs. Table \ref{tab:gpu_specs} lists the GPU models, along with the per-model number of GPUs, their idle power consumption (in Watt, mapped to $p_{idle}$ in Eq.\ref{eq: gpu pwr cons}), and their Thermal Design Power (TDP) representing the power consumption under maximum theoretical load. TDP, measured in Watt, is mapped to $p_{max}$ in Eq.\ref{eq: gpu pwr cons}. 
\begin{table}[h]
    \caption{GPU models in the trace data.}
    \label{tab:gpu_specs}    
    \centering
    \begin{tabular}{r|c|c|c}
        \textbf{GPU model} & \textbf{Amount} & \textbf{Power idle (W)} & \textbf{TDP (W)} \\
        \hline
        \textbf{V100M16} & 195 & 30 & 300 \\
        \textbf{V100M32} & 204 & 30 & 300 \\
        \textbf{P100} & 265 & 25 & 250 \\
        \textbf{T4} & 842 & 10 & 70 \\
        \textbf{A10} & 2 & 30 & 150 \\
        \textbf{G2 (A10)} & 4,392 & 30 & 150 \\
        \textbf{G3 (A100)} & 312 & 50 & 400 \\
        \bottomrule
    \end{tabular}
\end{table}
The datacenter includes two classified GPU models, 
G2 and G3. Nodes with G2 GPUs have 96 virtual CPUs and 393,216 MiBs of memory, while those with G3 GPUs have 128 virtual CPUs and 786,432 MiBs of memory. Based on these specifications, we assume G2 corresponds to A10 GPUs and G3 to A100 GPUs.
The trace data provides no information on CPU models. Therefore, we assume the datacenter uses the \textit{Intel Xeon ES-2682 V4}, a CPU model commonly found in Alibaba’s commercial instances. This CPU features a power consumption of 15W when idling (which we map to the $p_{idle}$ in Eq.\ref{eq: cpu pwr cons}) and a TDP of 120W (which we map to the $p_{max}$ in Eq.\ref{eq: cpu pwr cons}), and 16 physical cores (which we map to the function $ncores(\cdot)$ in Eq.\ref{eq: cpu pwr cons}).


\subsection{Evaluation Metrics}
\label{sec: exp metrics}

The metrics considered in the experiments are:
\begin{enumerate}
\item The \textbf{Estimated Overall Power Consumption} (\textbf{EOPC}) of the datacenter accounts for both CPU and GPU power consumption (in Watt). It is modeled as described in Equation \ref{eq: datacenter estimated power consumption} from Section \ref{section:background}. 
\item \textbf{GPU Resource Allocation Ratio (GRAR)}: the ratio between the sum of GPU resources allocated to scheduled tasks and the sum of GPU resources requested by arrived tasks. This measure is a proxy to assess how well each scheduling strategy manages GPU fragmentation (as defined in Equation \ref{eq: gpu frag} from Section \ref{section:background}) when the datacenter nears saturation. Low ratios indicate many task failures and thus poor task scheduling due to GPU fragmentation. 
\end{enumerate}
Each batch of experiments is repeated 10 times. For each metric, we report the average value relative to the cumulative GPU resource requests made by tasks arrived to the datacenter.

\section{Experimental evaluation}
\label{section:experimental eval}

The experimental evaluation is divided into four parts. First, in Section \ref{sec: exp fgd prw cons}, we provide an overview of the power consumption for the simulated GPU datacenter, establishing a baseline for later discussions on power savings. In Section \ref{subsec: comb pwr with fgd}, we assess the power savings of  \Nameapproach and its linear combinations with FGD, compared to plain FGD, focusing on minimizing both power consumption and GPU fragmentation. Section \ref{sec: broad eval pwr save} evaluates the power savings achieved by selected \Nameapproach and FGD combinations, as well as other competitors from Section \ref{sec: exp competitors}, relative to FGD. Finally, in Section \ref{sec: exp grar eval}, we evaluate the GRAR metric to confirm that the power savings are genuine and not due to task scheduling failures when the datacenter is not near saturation. We also show that the GRAR scores of the selected \Nameapproach and FGD combinations are close to those of plain FGD.

\subsection{The FGD EOPC baseline}
\label{sec: exp fgd prw cons}

\begin{figure}[t]
\centering
\includegraphics[width=.49\textwidth]{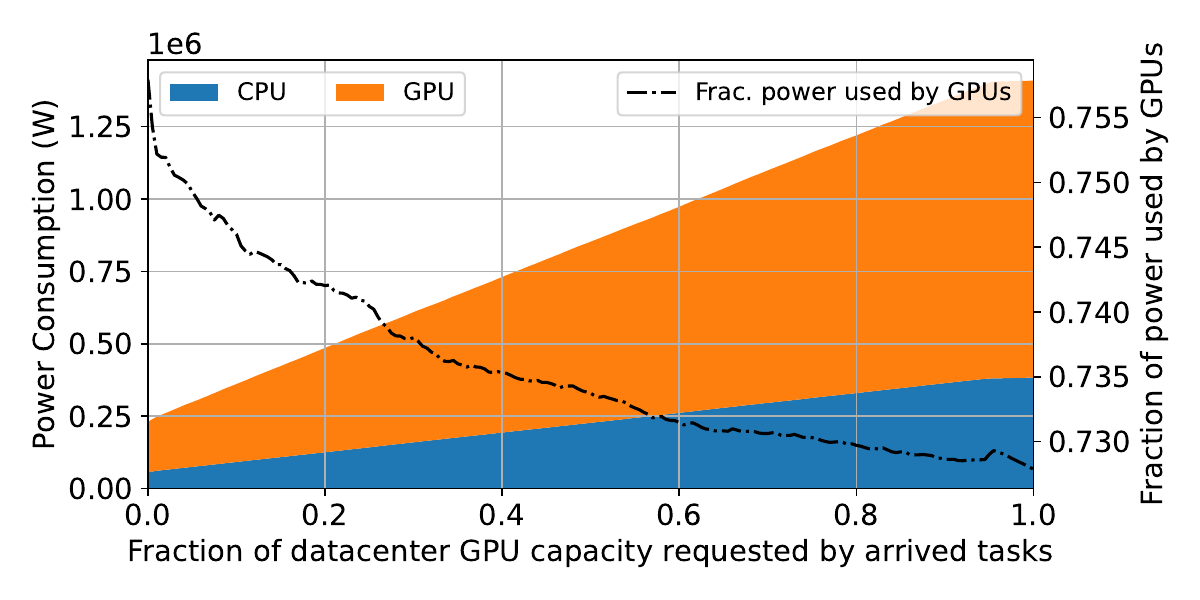}
\caption{FGD EOPC (in MW) for workloads from the Default trace, with stacked CPU and GPU components. The dashed line shows the fraction of GPU power (see right-hand y-axis).}
\label{fig: exp default fgd pwr cons}
\end{figure}

This first batch of experiments has the purpose of 
setting the stage for subsequent discussions on the power savings achieved by various competitors compared to FGD. To this end, we consider the EOPC achieved by FGD when considering workloads from the Default trace. The plot in Figure \ref{fig: exp default fgd pwr cons} presents the results, shown in the form of a stacked plot, where the CPU and GPU components of EOPC are highlighted.

From the figure, we observe that FGD EOPC starts just above 200kW and peaks at about 1.4MW as the simulated GPU datacenter approaches saturation. Additionally, by observing the dashed line associated with the right y-axis in the plot, we note that the estimated GPU power consumption consistently represents between 72\% and 76\% of the EOPC. Finally, we report that similar figures and trends are observed when estimating EOPC for workloads from the other traces.

\subsection{Combining \Nameapproach with FGD}
\label{subsec: comb pwr with fgd}

These experiments aim to identify the most effective linear combinations of \Nameapproach's and FGD's scores for minimizing both power consumption and GPU fragmentation -- recall that the Kubernetes scheduling framework allows the linear combination of scores from multiple scoring plugins. Following Equation \ref{eq: problem def} in Section \ref{section:background}, we denote such combinations by $\alpha \cdot \text{\Nameapproach} + (1-\alpha) \cdot \text{FGD}$, with $\alpha \in [0,1]$. 
Note that in all the plots, the values of $\alpha$ and $(1-\alpha)$ are shown multiplied by 1000.
To evaluate the combinations considered in the experiments, we focus on the two metrics introduced in Section \ref{sec: exp metrics}. We use EOPC to assess the power savings achieved by \Nameapproach and its combinations with FGD compared to plain FGD. At the same time, GRAR  helps us determine whether the power savings are genuine (i.e., not due to task scheduling failures) and identify which combinations minimize power consumption and GPU fragmentation. We report only the results of experiments conducted with workloads generated from the Default trace, as similar behaviors have been observed with the other traces.
\begin{figure}[t]
\centering
\includegraphics[width=0.49\textwidth]{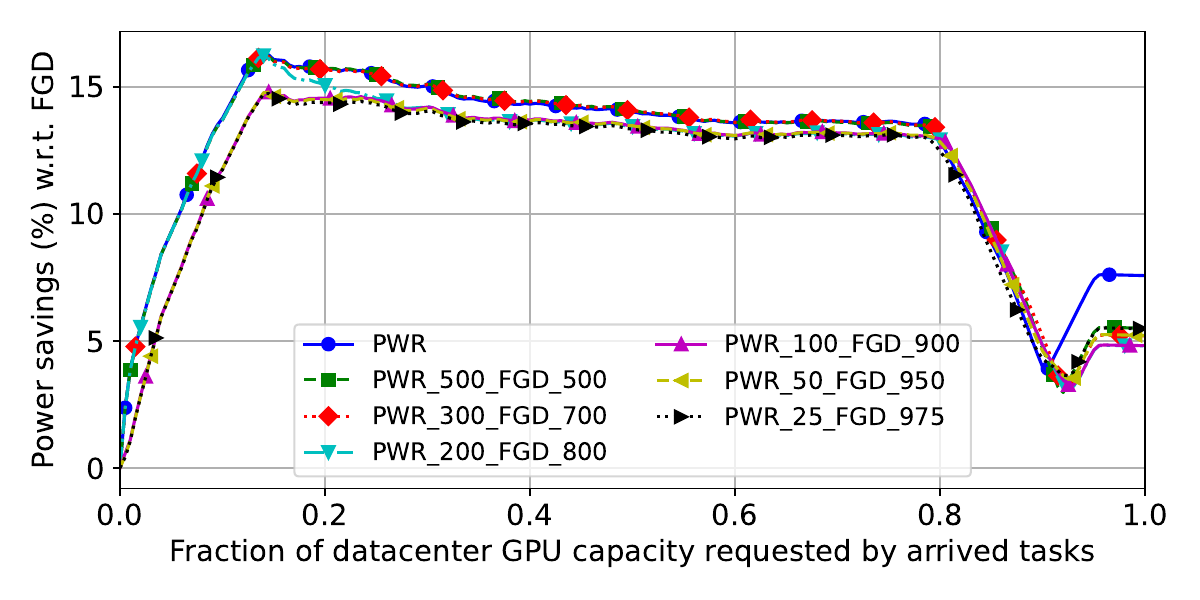}
\includegraphics[width=0.49\textwidth]{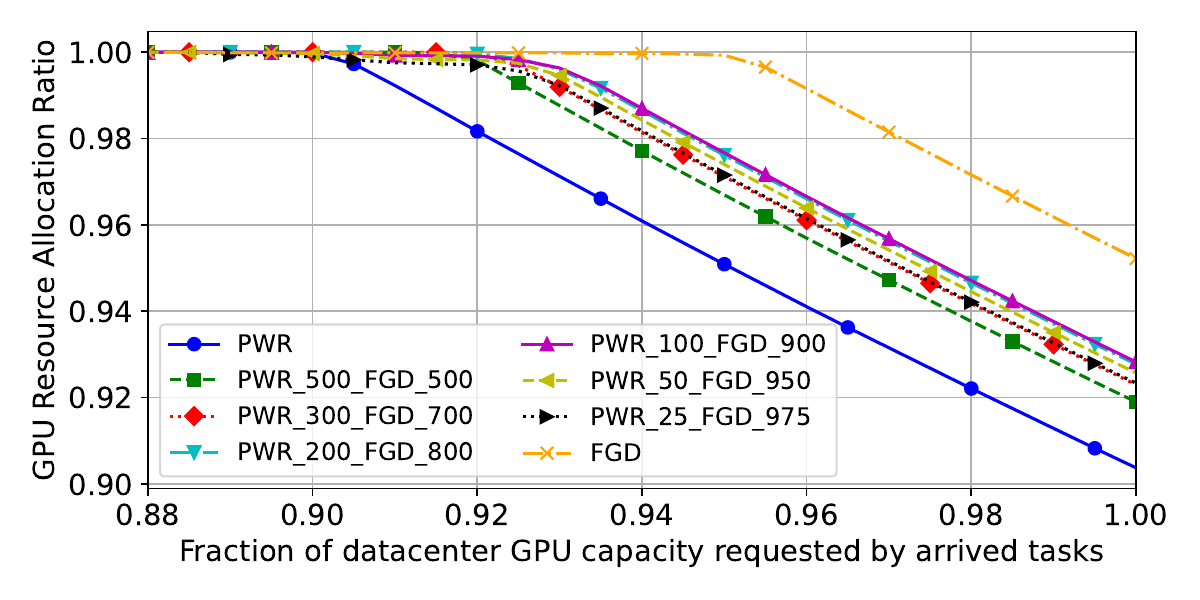}
\caption{Power savings (in percentage w.r.t. FGD, \textit{top plot}) and GRAR scores (\textit{bottom plot}) measured for \Nameapproach and its linear combinations with FGD, workloads from Default trace.}
\label{fig: exp special}
\end{figure}

We begin by examining the power savings achieved by \Nameapproach and its linear combinations with FGD compared to plain FGD (see Figure \ref{fig: exp special}, \textit{top plot}). Note that all the curves exhibit the same pattern, unfolding in five distinct phases: in phase (1) (x-axis interval $[0, 0.15]$), there is an \textbf{initial rapid gain} in power savings compared to FGD, followed in phase (2) (x-axis interval $[0.15, 0.8]$) by a \textbf{large region} of sustained power savings. Beyond this region, phase (3) (x-axis interval $[0.8, 0.9]$) shows a \textbf{steep decrease} in power savings, followed in phase (4) (x-axis interval $[0.9, 0.93]$) by a \textbf{small rebound}. Finally, in phase (5) (x-axis interval $[0.93, 1]$) there is a \textbf{final stabilization}.
As we will observe in experiments conducted in subsequent sections, this is a recurring pattern. Consequently, it is important to explain the underlying reasons behind these five phases. To this purpose, we examine the GRAR metric plotted in Figure \ref{fig: exp special}, \textit{bottom plot}. Note that the x-axis in the plot is zoomed in on the $[0.85,1]$ interval because, before the datacenter reaches $\sim$88\% of GPU capacity requested by arrived tasks, all competitors achieve a perfect GRAR score of 1, indicating no task scheduling failures up to that point. This, in turn, implies that in the region associated with phase (1), our solutions use fewer CPUs and GPUs -- hence less power -- than plain FGD to schedule the same number of tasks. Moreover, phase (2) shows that they maintain this advantage until the datacenter reaches approximately 80\% requested GPU capacity. 
The steep decreases in power savings vs FGD observed in phase (3) do not correspond to an immediate worsening of the GRAR metric. This indicates that the decreasing power savings are due to an increasing pressure on the dwindling idle resources as the datacenter approaches saturation. The small rebounds observed in phase (4) represent minor increases in power savings achieved by the competitors compared to plain FGD. Still, by looking at the GRAR metric, these increases are clearly due to an increasing number of failures of task scheduling requests compared to FGD. Finally, the stabilization observed in phase (5) occurs because, by that point, FGD also starts failing to schedule tasks.

We conclude by commenting on the performances of the various linear combinations of \Nameapproach and FGD tested.
By examining the power savings in Figure \ref{fig: exp special}, \textit{top plot}, we see two distinct groups: one consisting of solutions (i.e., those with $\alpha > 0.2$) whose power savings closely follow those of plain \Nameapproach, while the others (i.e., with $\alpha \leq 0.2$) achieve slightly lower power savings. Focusing on the GRAR metric (Figure \ref{fig: exp special}, \textit{bottom plot}), we see that those in the first group achieve slightly worse results than those in the second group. The combinations striking the best compromises are $\alpha \in \{0.05, 0.1, 0.2\}$. Intuitively, the more the weight given to FGD, the better the resulting GRAR tends to be, but the less the resulting power savings vs plain FGD; we report, however, that beyond $\alpha = 0.05$, the GRAR metric stops improving.
In summary, the more the weight given to \Nameapproach in the combination, the greater the power savings when the datacenter is not near saturation. However, this comes at the expense of earlier task scheduling failures compared to FGD, reflected in the lower GRAR scores, as the datacenter nears saturation.
Overall, the combinations of \Nameapproach and FGD striking the best compromises between power and GPU fragmentation minimization are those with $\alpha \in \{0.05, 0.1, 0.2\}$. In the following experiments, we focus on these three specific combinations.

\subsection{Comparison of power savings with competitors}
\label{sec: broad eval pwr save}

We evaluate the power savings achieved by the three combinations of \Nameapproach with FGD selected from Section \ref{subsec: comb pwr with fgd} and the competitors introduced in Section \ref{sec: exp competitors}. Power savings are measured using the EOPC metric w.r.t. plain FGD using workloads generated from the traces from Section \ref{sec: exp traces}.

\begin{figure}[t]
\centering
\includegraphics[width=.49\textwidth]{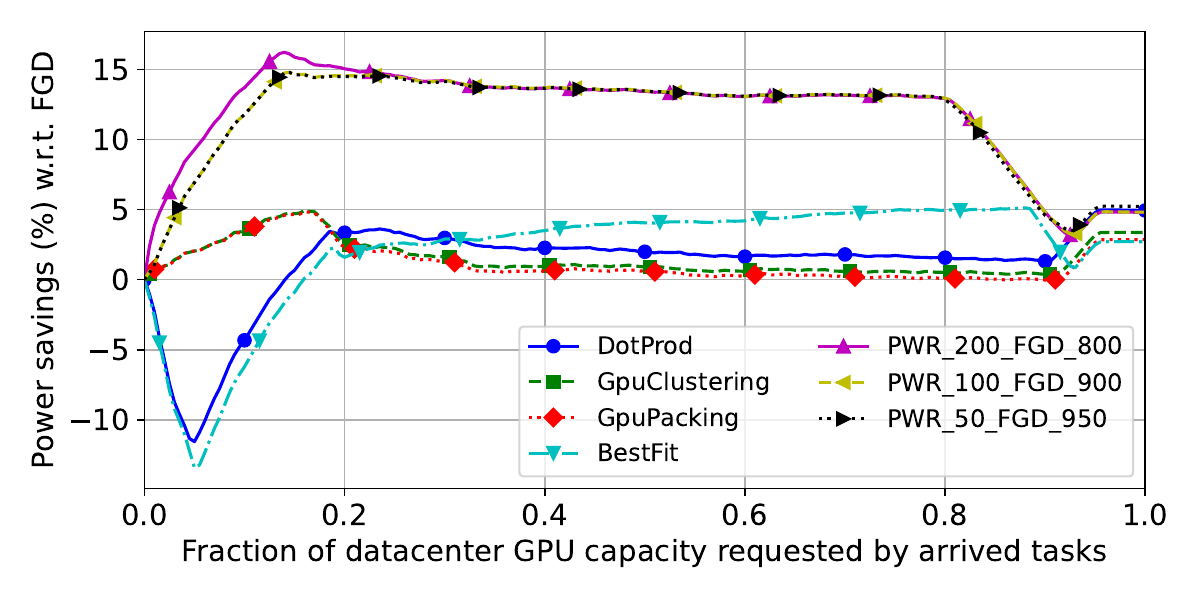}
\caption{Power savings with workloads from the Default trace.}
\label{fig: exp default pwr}
\end{figure}

\subsubsection{Workloads from the Default trace}
\label{sec:experimental eval pwr default}

The results of these experiments are presented in Figure \ref{fig: exp default pwr}.
Observe that our solutions consistently outperform the competitors over the whole capacity interval. They achieve power savings exceeding 13\% until the requested GPU capacity reaches $\sim$80\%, and then stay above $\sim$5\% until the requested GPU capacity reaches $\sim$90\%. 
Conversely, the competitors never exceed a 5\% power saving w.r.t. plain FGD. 


\subsubsection{Workloads from the sharing-GPU traces}
\label{sec:experimental eval pwr share-gpu}

\begin{figure}[t]
\centering
\includegraphics[width=.49\textwidth]{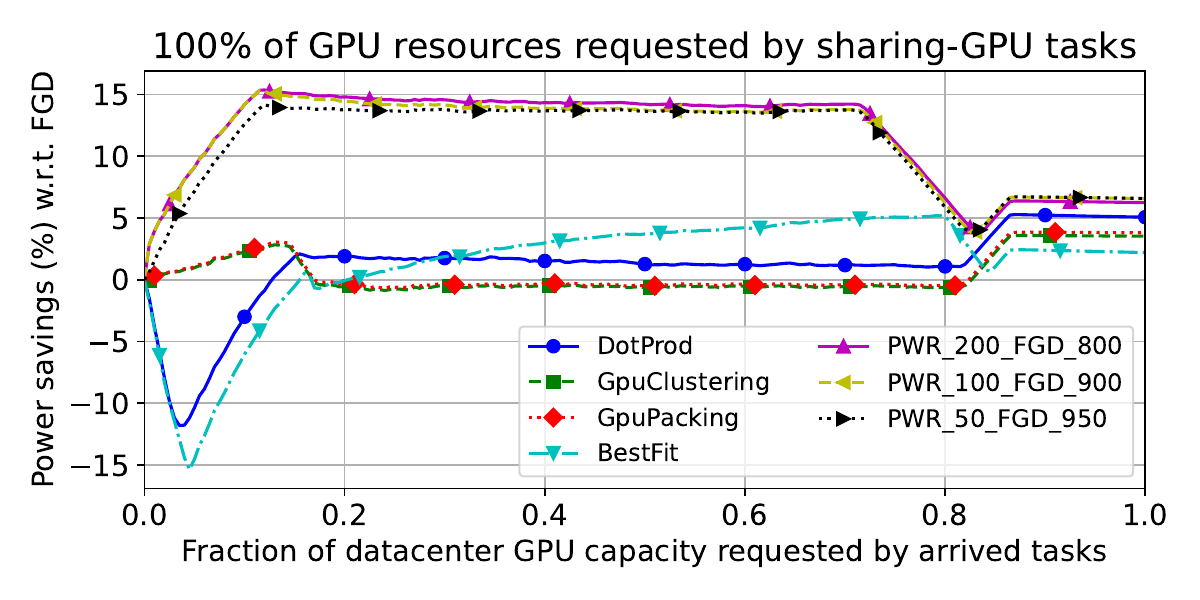}
\caption{Power savings with {sharing-GPU} workloads -- case in which sharing-GPU tasks request 100\% of GPU resources.}
\label{fig: exp sharegpu pwr}
\end{figure}

In these experiments, we consider workloads generated from the four sharing-GPU traces described in Section \ref{sec: exp traces}. The results are shown in Figure \ref{fig: exp sharegpu pwr}, where the plot focuses on the 100\% case (the trends observed with the other three cases are similar).
We observe a behavior similar to that discussed in Section \ref{sec:experimental eval pwr default}, including the five-phase pattern characterizing the curves of the three selected combinations of \Nameapproach with FGD. More specifically, these combinations achieve much larger power savings than competitors, with savings exceeding 13\% until the datacenter's requested GPU capacity reaches $\sim$70\%; subsequently, the savings stay above 5\% until the requested GPU capacity reaches $\sim$80\%.
The best competitor, BestFit, reaches the largest power saving of about 5\% only when the requested GPU capacity reaches $\sim$80\%.

\begin{figure}[t]
\centering

\includegraphics[width=0.49\textwidth]{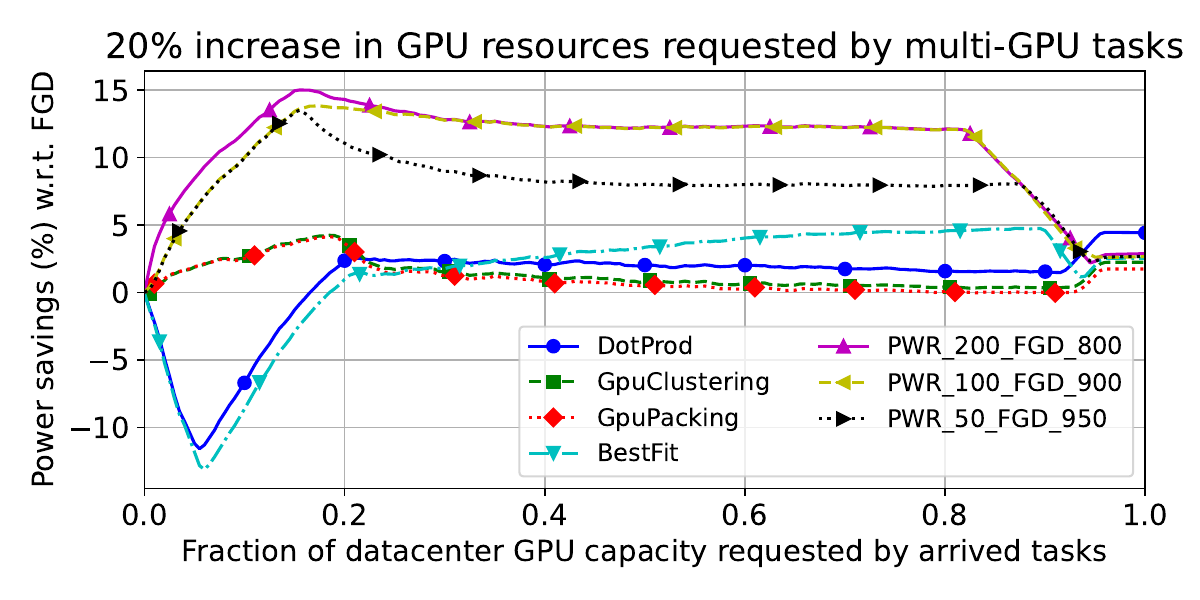}
\includegraphics[width=0.49\textwidth]{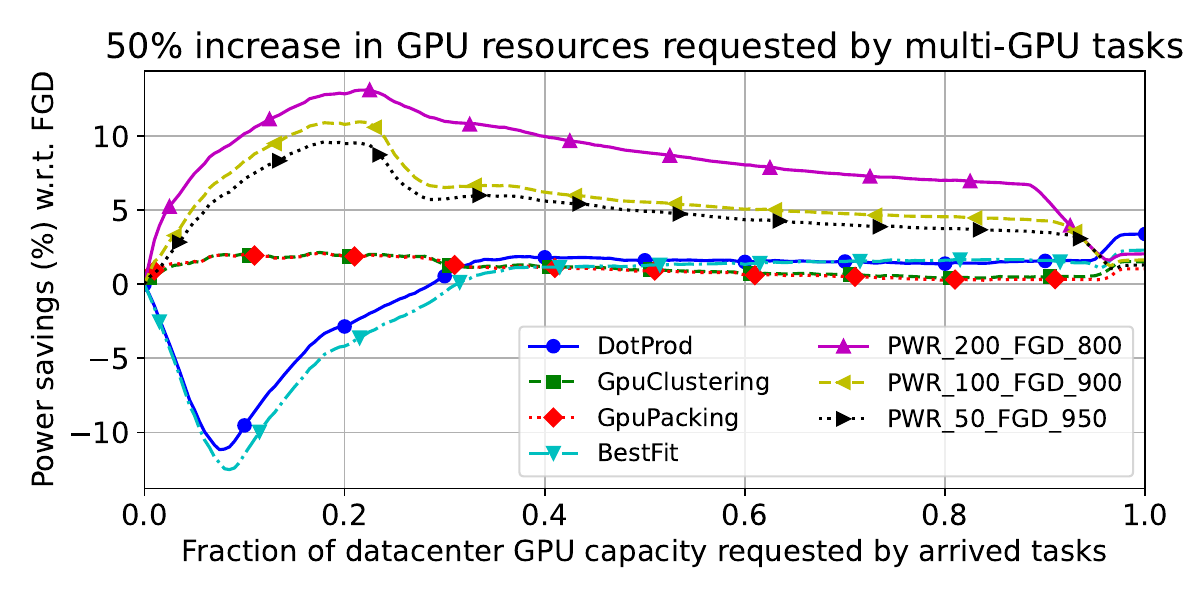}
\caption{Power savings with multi-GPU workloads.}
\label{fig: exp multigpu pwr}
\end{figure}

\subsubsection{Workloads from the multi-GPU traces}
\label{sec: experimental eval pwr multigpu}
In these experiments, we consider workloads generated from the four multi-GPU traces introduced in Section \ref{sec: exp traces}. The results are shown in the two plots of Figure \ref{fig: exp multigpu pwr}, and refer to the 20\% and 50\% cases (the other two intermediate cases fall between).
Again, the selected combinations of \Nameapproach with FGD largely achieve the best power savings over competitor methods. Specifically, in the 20\% case, the combinations with $\alpha \in \{0.1, 0.2\}$ consistently achieve power savings above 12\%, and the one with $\alpha = 0.05$ above 7\%, until the requested GPU capacity reaches $\sim$82\%. 
In the multi-GPU 50\% case, the power savings achieved by our three combinations are lower, but still relevant. Specifically, the combination with $\alpha = 0.2$ achieves power savings consistently above 7\% until the datacenter GPU requested capacity reaches $\sim$90\%, while the other two combinations achieve power savings consistently above 4\% in the same interval.
Finally, similarly to Sections \ref{sec:experimental eval pwr  default} and \ref{sec:experimental eval pwr share-gpu}, we still observe the five-phase pattern characterizing the curves of our solutions and the lower performance of all the competitors for requested capacities not close to saturation.

\subsubsection{Workloads from the GPU-constrained traces}

\begin{figure}[t]
\centering
\includegraphics[width=0.49\textwidth]{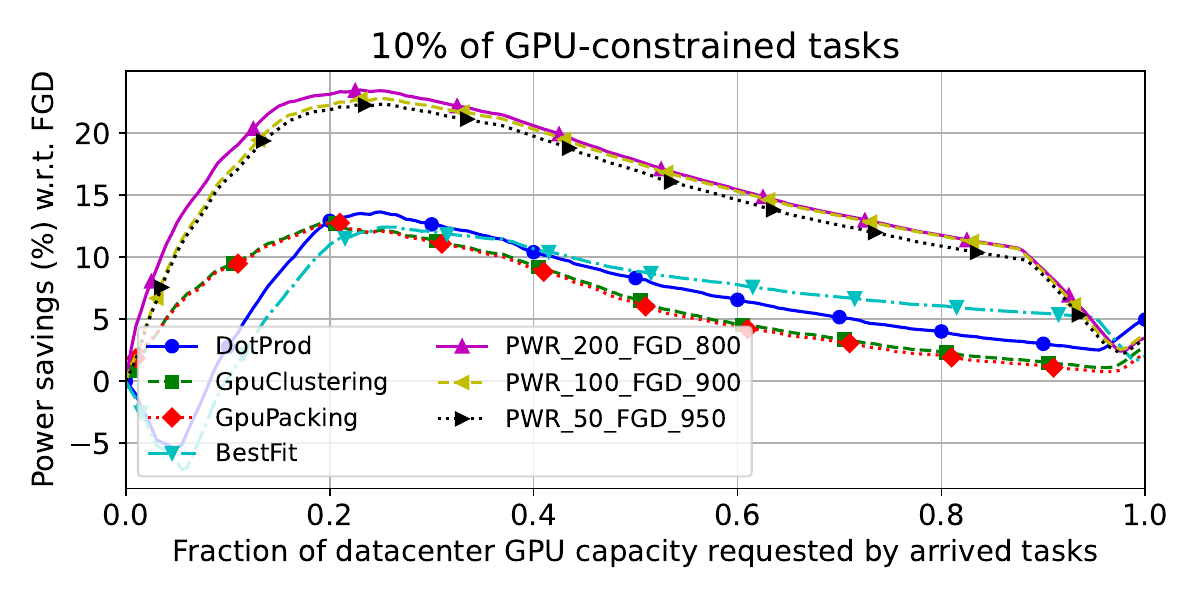}
\includegraphics[width=0.49\textwidth]{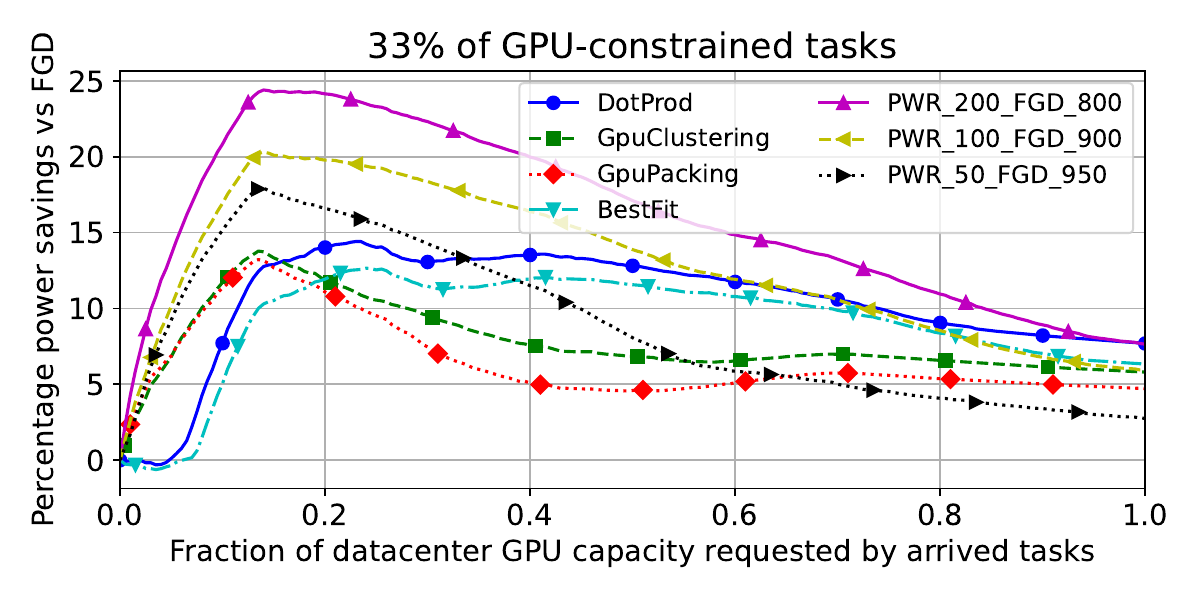}
\caption{Power savings with constrained-GPU workloads.}
\label{fig: exp constgpu pwr}
\end{figure}

Here we consider workloads generated from the four constrained-GPU traces introduced in Section \ref{sec: exp traces}. The results are shown in Figure \ref{fig: exp constgpu pwr}; note that the Figure shows the plots for the  10\% and 33\% cases, as the trends observed with the other two intermediate cases fall between these two.
On one end of the spectrum, we have the case in which GPU-constrained tasks represent the 10\% of the total GPU tasks. In this case, the three selected combinations still follow the five-phase pattern, and achieve large power savings consistently better than those of the competitors: power savings stay consistently above 10\% until the requested GPU capacity reaches $\sim$90\%. 
On the other end of the spectrum, i.e., the case in which GPU-constrained tasks represent the 33\% of the total, the curves referred to our three combinations appear to not follow the five-phase pattern anymore. Furthermore, the combination with $\alpha = 0.2$ still continues to achieve large power savings comparable to the 10\% case, but the other two combinations achieve lower power savings and do not consistently outperform some of the competitors -- this is especially true for the combination with $\alpha = 0.05$. 
We refer the reader to the analysis of the GRAR metric in Section \ref{sec: experimental eval gpuocc constgpu} for more insights on the results with this type of trace. 

\subsection{Evaluation with the GRAR metric}
\label{sec: exp grar eval}


This section evaluates the GRAR metric achieved by the three selected combinations and their competitors.
As a GPU datacenter approaches saturation, a policy that effectively minimizes GPU fragmentation -- and thus optimizes GRAR -- will be able to schedule more tasks using the same amount of GPU resources.
Furthermore, showing that a scheduling policy can achieve GRAR scores comparable to other policies while also exhibiting significant power savings would indicate that the savings are not due to failures in scheduling tasks, which would otherwise leave the datacenter's resources unused or underutilized.
Specifically, we aim to determine (1) to what extent our three selected combinations of \Nameapproach with FGD can yield perfect GRAR scores, thus proving that the power savings they achieve are real in the corresponding capacity intervals. Furthermore, we aim to determine whether (2) our solutions can achieve GRAR scores reasonably close to those obtained by plain FGD, (3) these combinations can ideally yield GRAR scores that are better or on par with BestFit (which has been identified as the second-best scheduling policy for reducing GPU fragmentation in \cite{weng2023beware}), and (4) they consistently outperform the remaining competitors.

\begin{figure}[t]
\centering
\includegraphics[width=.49\textwidth]{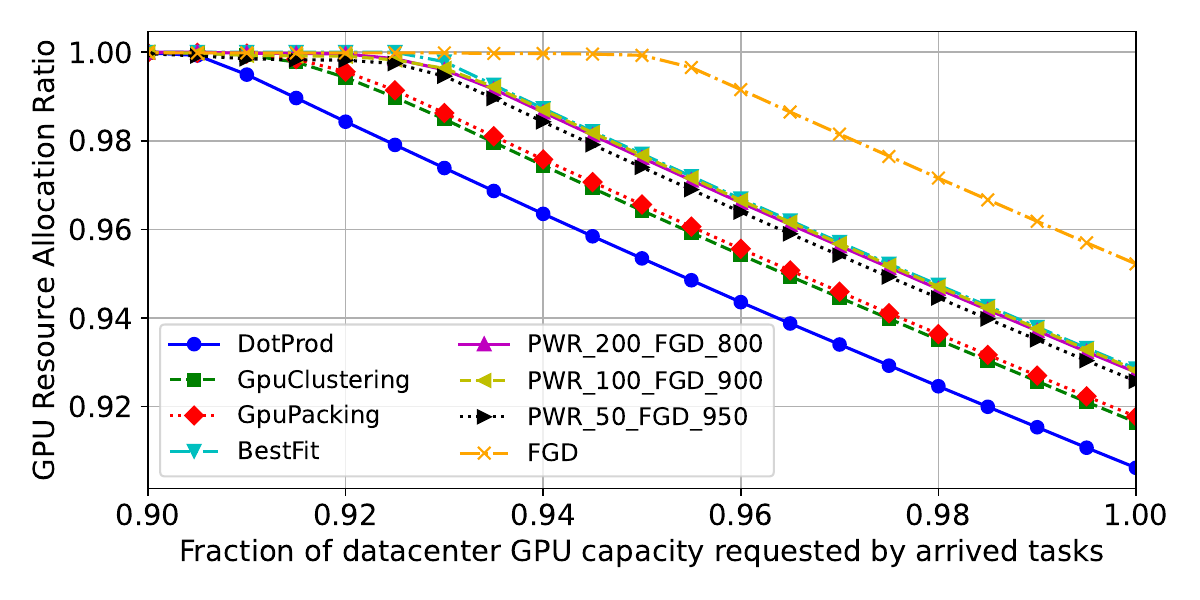}
\caption{GRAR with workloads from the Default trace.}
\label{fig: exp default gpuocc}
\end{figure}

\begin{figure}[t]
\centering
\includegraphics[width=.49\textwidth]{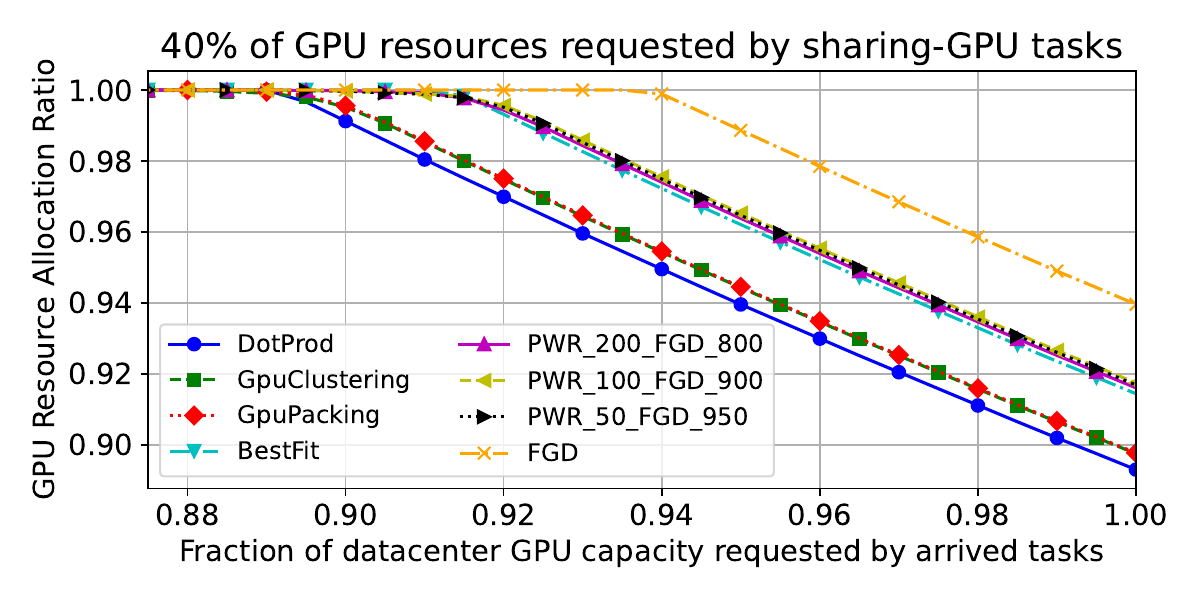}
\includegraphics[width=.49\textwidth]{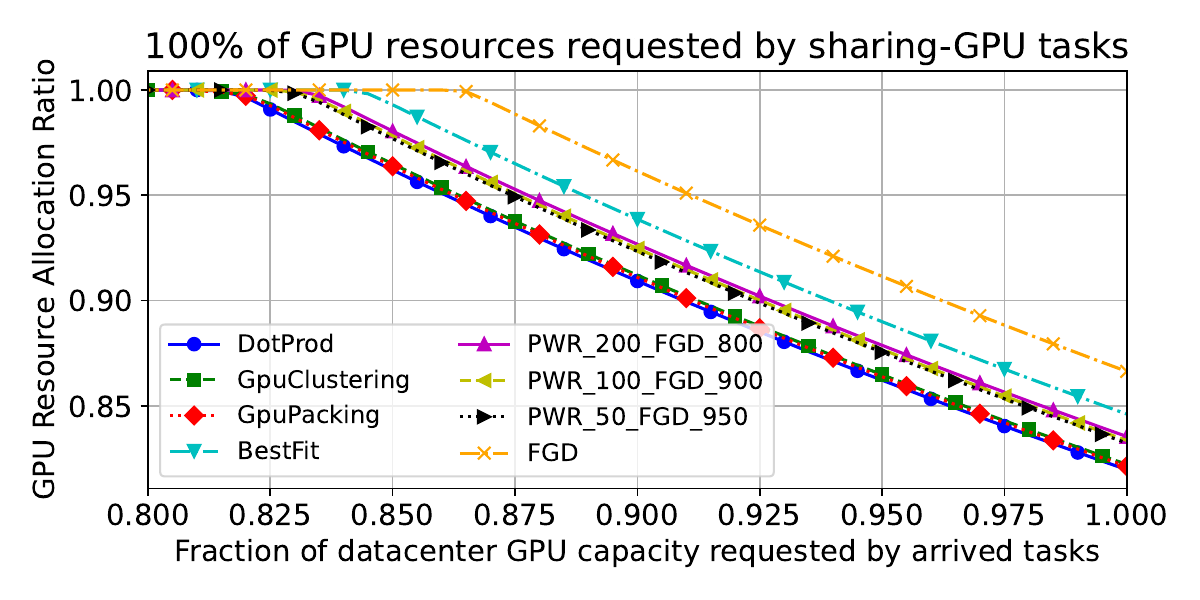}
\caption{GRAR with sharing-GPU workloads.}
\label{fig: exp sharegpu gpuocc}
\end{figure}

\subsubsection{Workloads generated from the Default trace}

The results of these experiments, conducted with workloads generated from the Default trace, are shown in Figure \ref{fig: exp default gpuocc}. Note that the x-axis in the plot is zoomed in on the [0.9, 1] interval, since the GRAR metric starts to worsen for all the competitors only when the datacenter GPU capacity requested by arrived tasks goes beyond 90\%.
Observing the plot, we see that as expected plain FGD achieves the best GRAR scores, closely followed by a group made of BestFit and our combinations of \Nameapproach with FGD (gap of around 2\% from plain FGD). The remaining competitors, then, close the rank (GRAR gap around 0.01).
Overall, the GRAR metric starts to worsen for all the considered competitors only when 
the datacenter GPU capacity requested by arrived tasks goes beyond 85\%: this implies that the power savings achieved by the three selected combinations, as shown in Section \ref{sec:experimental eval pwr default}, are not due to task scheduling failures.

\subsubsection{Workloads from the sharing-GPU traces}

Let us now consider workloads generated from the four \textit{sharing-GPU} traces introduced in Section \ref{sec: exp traces}. The results of these experiments are shown in Figure \ref{fig: exp sharegpu gpuocc}, where the plots focus on the 40\% and 100\% cases -- the results observed with workloads from the other two intermediate traces fall between these two. Note that the x-axis in the plots is zoomed, since the GRAR metric worsens for all scheduling policies only when the datacenter GPU capacity requested by arrived tasks goes beyond 80\%.

In the 40\% case, we see that plain FGD achieves the best GRAR scores, closely followed by a group made of BestFit and the three selected combinations (GRAR gap around 0.02). Finally, the remaining competitors closely follow the above group (GRAR gap 0.02).
As we increase the fraction of GPU resources requested by sharing-GPU tasks, we notice that BestFit performs slightly better than the three combinations. Indeed, in the 100\% case, BestFit follows plain FGD (GRAR gap around 0.02), which is then followed by a group made of the three combinations (GRAR gap around 0.01). The remaining competitors close the rank (GRAR gap around 0.01).
In all cases, the GRAR metric starts to get worse for all the considered competitors only when the datacenter GPU capacity requested by arrived tasks goes beyond 80\%: once again, this implies that the power savings achieved by the combinations of \Nameapproach and FGD shown in Section \ref{sec:experimental eval pwr share-gpu} are not due to task scheduling failures.

\subsubsection{Workloads from the multi-GPU traces}

\begin{figure}[t]
\centering
\includegraphics[width=0.49\textwidth]{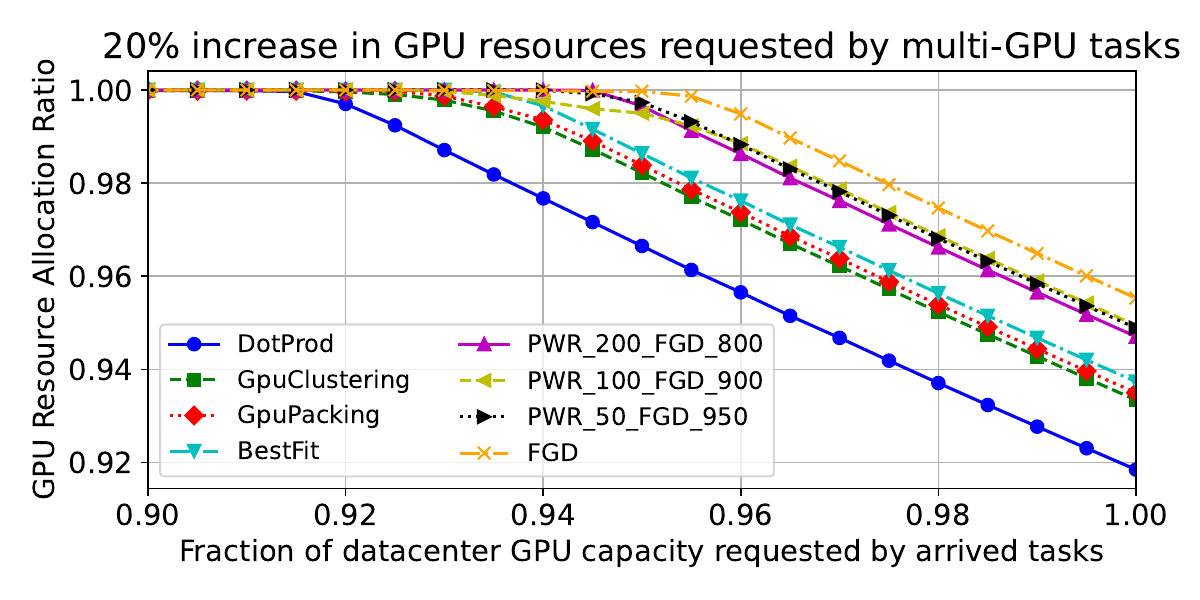}
\includegraphics[width=0.49\textwidth]{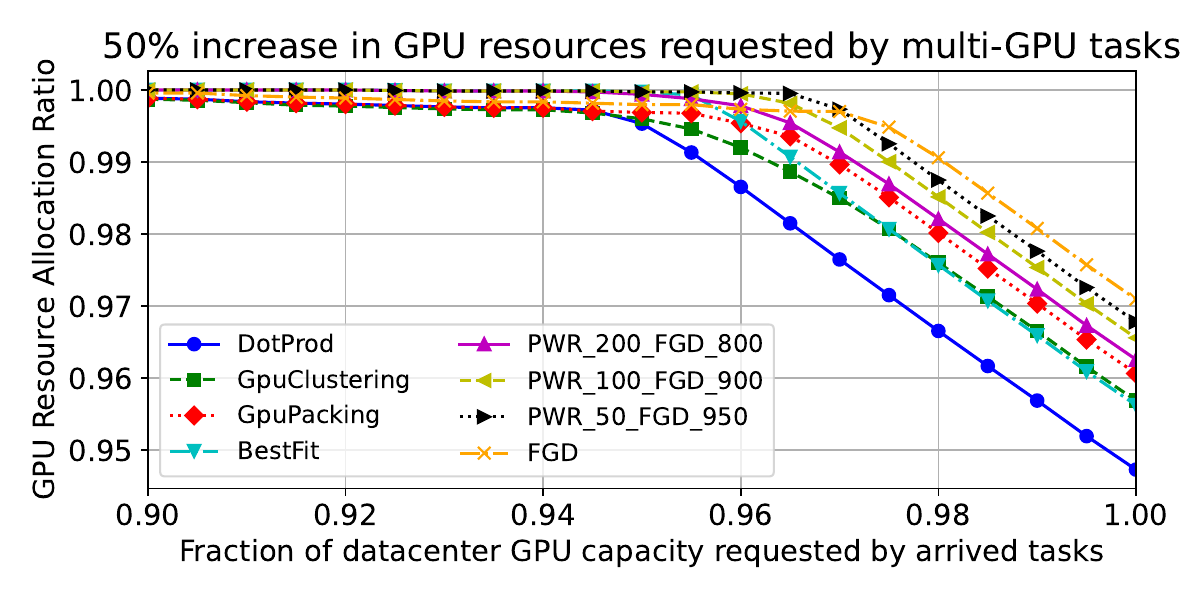}
\caption{GRAR with multi-GPU workloads.}
\label{fig: exp multigpu gpuocc}
\end{figure}

We analyze workloads generated from the four multi-GPU traces introduced in Section \ref{sec: exp traces}. Figure \ref{fig: exp multigpu gpuocc} summarizes the results, focusing on the 20\% and 50\% cases as the other two intermediate cases fall between.
The x-axis in the plots is zoomed to the [0.9, 1] range, as the GRAR metric starts to degrade noticeably only when GPU demand exceeds 90\% of the datacenter capacity.
For the 20\% case, plain FGD achieves the best GRAR, followed closely by the three combinations (with a GRAR difference of less than 0.01). BestFit and two other competitors follow (with a GRAR gap of around 0.01), while DotProd ranks last. As the fraction of GPU resources requested by multi-GPU tasks further increases, the gaps between competitors narrow, though the ranking remains the same. GRAR noticeably declines only when GPU demand exceeds 85\%, confirming that combining \Nameapproach with FGD provides practical power savings also under these workloads.

\subsubsection{Workloads from the GPU-constrained traces}
\label{sec: experimental eval gpuocc constgpu}

\begin{figure}[t]
\centering
\includegraphics[width=0.49\textwidth]{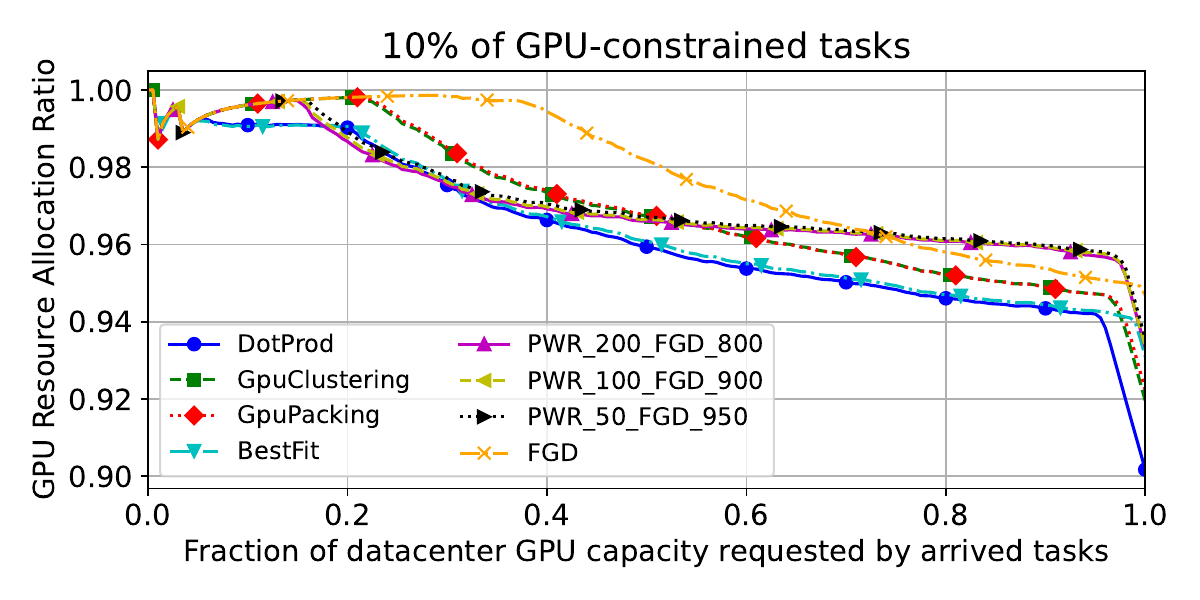}
\includegraphics[width=0.49\textwidth]{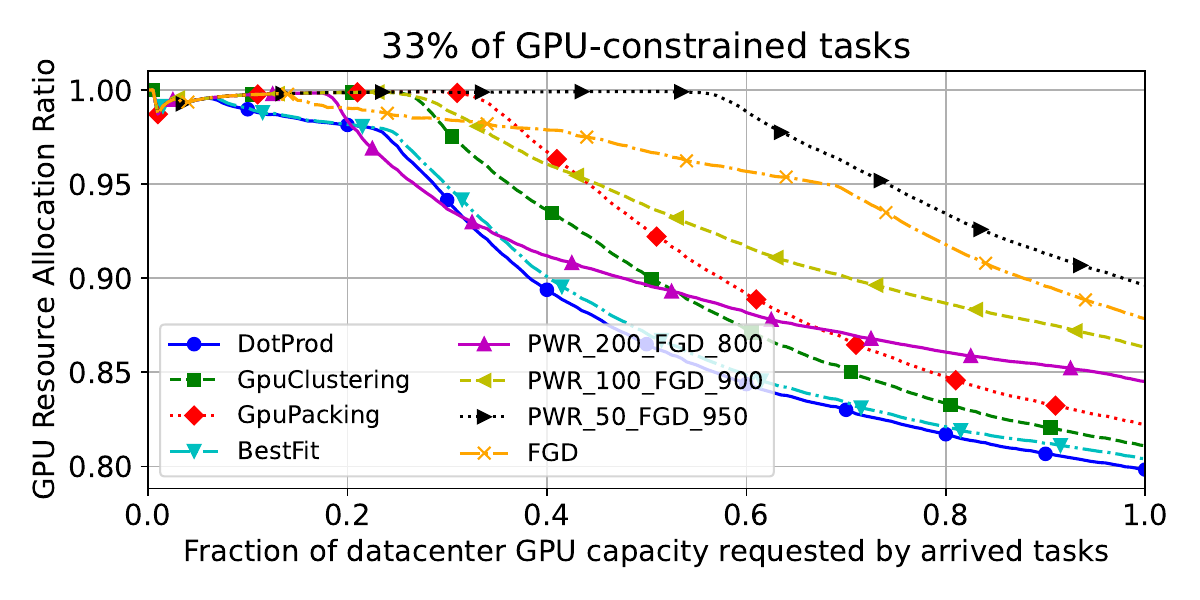}
\caption{GRAR with GPU-constrained workloads.}
\label{fig: exp constgpu gpuocc}
\end{figure}

In this set of experiments, we analyze workloads generated from the four GPU-constrained traces introduced in Section \ref{sec: exp traces}. Figure \ref{fig: exp constgpu gpuocc} focuses on the cases where GPU-constrained tasks represent 10\% and 33\% of GPU tasks, with results for the other two intermediate cases falling between.
In both plots, all schedulers  struggle to schedule tasks already before GPU demand reaches 80\% of the capacity. This is due to the pressure on specific GPU models caused by task constraints. While FGD and our three combinations achieve the best GRAR scores when GPU demand reaches 100\%, performance varies more significantly when the datacenter is not fully saturated, particularly in the 33\% case.
The final GRAR score is key for assessing how well a scheduling policy utilizes GPU resources. However, to understand power savings, it is important to analyze the GRAR differences before saturation, as some savings may result from task scheduling failures rather than efficient resource use.
In the 10\% case, plain FGD shows a maximum GRAR advantage of $\sim$0.025 over the three combinations in the [0.21, 0.73] GPU capacity range. This small gap is consistent with previous experiments, and further highlights the significant power savings achieved by our solution.
In the 33\% case, \Nameapproach with $\alpha = 0.05$ consistently outperforms plain FGD with a maximum GRAR gap of $\sim$0.04, explaining its smaller power savings w.r.t. the other two combinations and some competitors in certain GPU capacity sub-intervals. Finally, the combinations with $\alpha \in \{0.1, 0.2\}$ perform worse than plain FGD, with maximum GRAR gaps of $\sim$0.05 and $\sim$0.08, respectively, which accounts for their larger power savings.

\section{Conclusions}
\label{sec: conclusions} 

In this paper, we addressed the online scheduling problem in GPU datacenters, reducing power consumption and minimizing GPU fragmentation. 
We introduced \Nameapproach, a power-aware scheduling policy that optimizes power usage by selecting power-efficient GPU and CPU combinations. This is achieved through a simplified power consumption model integrated into a new Kubernetes score plugin. By combining our policy with the Fragmentation Gradient Descent (FGD) one, we achieve a balanced tradeoff between reducing power consumption and minimizing GPU fragmentation.
Extensive experiments, conducted using traces with different characteristics from the 2023 Alibaba GPU trace dataset, show that our approach offers substantial power savings compared to FGD alone. Specifically, the power savings vs plain FGD exceed 20\% for moderate workloads while maintaining efficient GPU utilization as a datacenter nears saturation. 
Our approach is a scalable and practical solution for future GPU datacenters, where usage and power efficiency will be both essential. In future work, we plan to refine the power model for greater accuracy, and possibly extend the approach to other resource-intensive environments. 
Another research direction involves studying under which conditions dynamically adjusting the coefficient $\alpha$ can improve power savings and GPU fragmentation.
%
%
Furthermore, we plan to integrate the notion of target workload into \Nameapproach to estimate the expected increase in power consumption when scheduling tasks, 
to further increase power savings.
Finally, we aim to extend our approach to batch scheduling, where optimizing across groups of tasks can further benefit power savings and GPU fragmentation. 
This extension could also
consider factors such as carbon footprint and energy costs into the optimization problem, potentially contributing to a more holistic approach to sustainability in GPU datacenters.

\section*{Acknowledgments} 

Funding for this research has been provided by Spoke 1 "Human-centered AI" of the M4C2 - Investimento 1.3, Partenariato Esteso PE00000013 - "FAIR - Future Artificial Intelligence Research". However, the views and opinions expressed are those of the authors only and do not necessarily reflect those of the EU or European Commission-EU. Neither the EU nor the granting authority can be held responsible for them.

\bibliographystyle{abbrv}
\bibliography{biblio}

\end{document}